\documentstyle[preprint,aps,eqsecnum]{revtex}
\tighten
\begin{document}
\preprint{EHU-FT/9602, hep-th/9611143}
\draft
\title{The Renormalization Group method for simple operator problems
in quantum mechanics}
\author{I.L.
Egusquiza\cite{emUPV} and M.A. Valle Basagoiti\cite{emmanu}}
\address{Dept. of Theoretical Physics\\
University of the Basque Country\\
 644 P.K. - 48080 BILBAO\\
SPAIN}
\date{November 18, 1996}
\maketitle
\begin{abstract}
A simple backreaction problem in quantum mechanics, the full quantum
anharmonic oscillator, and quantum parametric resonance are studied using 
Renormalization Group techniques for global asymptotic analysis. In this short
note this technique is adapted for the  first time to operator problems.
\end{abstract}

\section{Introduction and Motivations}

In many different contexts perturbation theory fails miserably because of the
growth of higher order terms, contrary to the basic perturbative assumption.
This secularity is present in both classical and quantum theories, and pervades
the motivation for the search for analytical methods to improve on perturbative
expansions.

There is a set of problems of particular interest to us where improvement on
perturbative expansions has shown itself to be of immediate necessity, and of
direct physical meaning, namely, (p)reheating in inflationary
cosmology \cite{yoshi,bts,linde,daniel}. We shall not deal directly with this
problem, though, and we will content ourselves with an analysis of two simple
quantum mechanical problems. The choice of problems and treatment thereof,
however, will be inspired by the mentioned physical setting.

The method we use is novel, and subsumes many other previously known ones. It
is the Renormalization Group (RG) method for global asymptotic analysis, as
advocated by Goldenfeld and collaborators \cite{goldetal,golddos}, which we
extend to operator problems in this short note. The key idea of the RG method
for global asymptotic analysis is the introduction of a time
parameter, additional to the initial value point, in such a way that the
perturbation expansion is valid in the vicinity of the introduced time
parameter. The coupling constants/constants of motion/initial conditions
(depending on your viewpoint and background one or another of these descriptions
will be more suitable) are turned into running constants, that is to say that
these constants are suitably modified by the change of the introduced time
parameter. On the other hand, the solution itself cannot depend on the
additional, new time parameter, so derivation with respect to the latter of the
perturbative solution will impose evolution equations for the running
constants. These equations are then solved for the running constants, and on
substitution in the perturbative expansion, together with the choice that the
time parameter is constantly updated to be time itself, we obtain an improved
solution.

This method has the clear advantage over  multiple scales perturbation analysis
\cite{benor,smith} that no a priori determination of the scales that appear in
the problem is necessary, and a naive perturbation expansion is enough as a
starting point. Many examples and illustrations of this advantage of the RG
method can be found in the works of Goldenfeld and collaborators.

The multiple scales method itself has been applied to operator problems
\cite{bb,frasca}, but not the RG method, and this paper is the first example,
to the best of our knowledge, of the application of the RG method to operator
problems. We shall choose the quantum anharmonic oscillator and the phenomenon
of quantum parametric resonance as our case studies, because of the important
r\^ole they have traditionally had as theoretical laboratories for new
perturbative methods, and because of their paradigmatic character in the
context of cosmological (p)reheating. Our results are comparable to all other
methods, and since they are obtained with a modest effort, we think that the
RG method is highly competitive in the operator context as well. There are a
number of areas where its usefulness might be proved, such as quantum optics,
but that we leave for further work.

\section{Setting of the problem}
The problem we shall first address is the back-reaction problem of the quantum
anharmonic oscillator; that is to say, the problem posed by the backreaction of
quantum fluctuations on the expected value of the position. Consider thus a
lagrangian of the form
\begin{equation}
L=\frac12 m\dot q^2 -\frac12 m\omega^2 q^2 -\frac14\lambda m q^4\,.
\end{equation}
The equation of motion in Heisenberg's picture reads
\begin{equation}
{{{\rm d}^2q}\over{{\rm d}t^2}} + \omega^2 q + \lambda q^3=0\,,\label{heiseq}
\end{equation}
which is, of course, an operator equation, the quantum Duffing equation. Let us
now decompose the
$q$ operator into a c-number times the unit operator and a fluctuation operator,
in such a way that the expectation value of $q$ is equal to the mentioned
c-number:
\begin{equation}
q=\varphi+\xi\,,
\end{equation}
where $\varphi=\langle q\rangle$, $\langle\xi\rangle=0$, the unit operator is
omitted, and $\langle\cdot\rangle$ stands for the expectation value of an
operator over a given state, which, as we are in Heisenberg's picture, is a
time independent state, and carries the initial value data for the operator
evolution equation. Both the c-number $\varphi$ and the operator $\xi$ are time
dependent, of course.

Taking the expectation value of the operator equation (\ref{heiseq}) we obtain
\begin{equation}
{{{\rm d}^2\varphi}\over{{\rm d}t^2}} + \omega^2 \varphi +
\lambda\left(\varphi^3 + 3\varphi\langle\xi^2\rangle + \langle\xi^3\rangle
\right)=0\,,\label{fullexeq}
\end{equation}
and using this result, together with (\ref{heiseq}), we can write
\begin{equation}
{{{\rm d}^2\xi}\over{{\rm d}t^2}} + \left(\omega^2 +3\lambda\varphi^2\right)\xi =
3\lambda\varphi\left(\langle\xi^2\rangle-\xi^2\right)
+\lambda\left(\langle\xi^3\rangle-\xi^3\right)\,.\label{fullfluceq}
\end{equation}
It is clear that the pair of equations (\ref{fullexeq}) and (\ref{fullfluceq}) is
equivalent to the operator equation (\ref{heiseq}) together with an initial
condition provided by the state over which the expectation values are computed.

So far we have followed the computations in Ehrenfest's theorem \cite{galindo},
and the result is, as is well known in elementary quantum mechanics, that the
evolution of the expectation value of the $q$ operator is not given by the
classical evolution equations: there are quantum corrections because the
potential is not quadratic. 

The customary way of handling this kind of problem is to revert to the
interaction picture and use time independent perturbation theory. We shall
depart from this well-trodden route and study the large time asymptotics of the
expectation value from a truncation of the exact equations (\ref{fullexeq}) and
(\ref{fullfluceq}). The truncation, which we will call the one-loop approximation
because of the analogy to the one-loop truncation in quantum field theory,
leaves us the following coupled equations, one  c-number and the other 
operator valued:
\begin{eqnarray}
{{{\rm d}^2\varphi}\over{{\rm d}t^2}} + \left(\omega^2 +
3\lambda\langle\xi^2\rangle\right)\varphi & = & -\lambda\varphi^3\,,\nonumber\\
{{{\rm d}^2\xi}\over{{\rm d}t^2}} + \left(\omega^2 +3\lambda\varphi^2\right)\xi &
= & 0\,.\label{oneloop}
\end{eqnarray}
Notice that for these truncated equations the equation for fluctuations is
linear, with a time-dependent frequency related to $\varphi^2$. We could now
try a na\"\i ve perturbation expansion for these equations, of the form
\begin{eqnarray}
\varphi & = & \varphi_0 +\lambda\varphi_1 +O(\lambda^2)\,,\nonumber\\
\xi& = & \xi_0 +\lambda\xi_1 +O(\lambda^2)\,,\nonumber
\end{eqnarray}
which, when inserted in eqns (\ref{oneloop}), yields
\begin{eqnarray}
{{{\rm d}^2\varphi_0}\over{{\rm d}t^2}} + \omega^2\varphi_0 & = &
0\,,\nonumber\\
{{{\rm d}^2\varphi_1}\over{{\rm d}t^2}} + \omega^2\varphi_1 & = &
-3\langle\xi_0^2\rangle\varphi_0 - \varphi_0^3\,,\nonumber\\
{{{\rm d}^2\xi_0}\over{{\rm d}t^2}} + \omega^2\xi_0 & = &
0\,,\nonumber\\
{{{\rm d}^2\xi_1}\over{{\rm d}t^2}} + \omega^2\xi_1 & = & -3\varphi_0^2\xi_0\,,
\label{pertur}
\end{eqnarray}
etc. It now becomes apparent that even within this simple truncation the
perturbative expansion cannot work, since secular terms will appear and the
perturbative hypothesis, namely, that the terms in higher order in $\lambda$
are smaller than the previous ones, will stop holding after a time of the order
$1/\lambda$. This is the issue we shall now address by means of the RG method.
\section{Application of the RG method}
The zeroth order equations in the series (\ref{pertur}) are solved by
\begin{eqnarray}
\varphi_0 & = & R\cos(\omega t +\theta)\,,\nonumber\\
\xi_0 & = & l\left(\alpha^{\dag}  e^{i\omega t} + \alpha e^{-i\omega
t}\right)\,,\label{zeroth}
\end{eqnarray}
where $\alpha$ and $\alpha^{\dag}$ are formally adjoint to each other. The
commutation relations among these two operators are fixed by imposing canonical
commutation relations for $q$ and its conjugate operator $p$. We obtain
\begin{equation}
[\alpha,\alpha^{\dag}]= {{\hbar}\over{2 m \omega l^2}}\,,
\end{equation}
and, therefore, choosing $l=\sqrt{\hbar/(2m\omega)}$, $\alpha$ and
$\alpha^{\dag}$ can be understood as annihilation and creation operators for a
harmonic oscillator, implying that we have a way of defining a number operator
$N=\alpha^{\dag}\alpha$ that will provide us with information about the
magnitude of the fluctuations.

So far everything has been straightforward. Problems first arise in computing
the first order corrections, since secular terms will be generated. Consider a
time $\tau$ for which the first order corrections vanish. The solutions to the
first order equations will be
\begin{eqnarray}
\varphi_1 &=& {{3 i R}\over{16\omega}}(t-\tau)\times
\Bigg(e^{i\omega
t}\left(R^2 e^{i\theta} + 4 l^2\langle\left(\alpha^{\dag}\right)^2\rangle
e^{-i\theta} +
 8 l^2\left(\langle N\rangle +\frac12\right)
e^{i\theta}\right)-\nonumber\\
& - &
e^{-i\omega t}\left(R^2 e^{-i\theta} + 4
l^2\langle\alpha^2\rangle e^{i\theta} + 8 l^2\left(\langle N\rangle
+\frac12\right) e^{-i\theta}\right)\Bigg)
 + {\rm r.t.}\,,
\end{eqnarray}
and
\begin{equation}
\xi_1 = {{3 i l R^2}\over{8\omega}}(t-\tau)
 \left(e^{i\omega
t}\left(2\alpha^{\dag} + \alpha e^{2i\theta}\right) - e^{-i\omega
t}\left(2\alpha + \alpha^{\dag} e^{-2i\theta}\right)\right) +{\rm r.t.}\,,
\end{equation}
where r.t. stands for regular terms, that is to say those which are bounded
when $t-\tau$  tends to infinity.

It is clear that for small $t-\tau$ the perturbative expansion has no {\sl a
priori} reason not to hold. Now, the crucial point is to realize that the
choice of $\tau$ corresponds to a choice of initial time for perturbations,
which, in other words, means that a particular value for the magnitudes $R$ and
$\theta$ and a particular choice for the (constant) operators $\alpha$ and
$\alpha^{\dag}$  has been chosen. If we were to allow those to change with
$\tau$, we would be able to readjust the perturbative series as we went away
from a given $\tau$ for a fixed $t$ to a different $\tau'$, around which the
perturbative expansion for $t-\tau'$ would now proceed. The change in $R$,
$\theta$, $\alpha$ and $\alpha^{\dag}$ would compensate the change in $\tau$.

On the other hand, the solution to the  differential equations which we intend
to solve by a perturbative expansion cannot depend on $\tau$. Once initial
values $R$ and $\theta$ and initial operators $\alpha$ and $\alpha^{\dag}$ are
given, the evolution of $\varphi$ and $\xi$ is fixed. Therefore, we obtain the
first order RG equations by allowing the stated dependence of the
(number and operator) parameters in $\tau$, deriving the expressions
$\varphi_0+\lambda\varphi_1$ and $\xi_0+\lambda\xi_1$ with respect to $\tau$,
and setting those derivatives equal to zero. To keep consistency with the
perturbative expansion, the terms of the form $\lambda\partial R/\partial\tau$
and similar are discarded, since they are of order $\lambda^2$, which is being
disregarded. 

Reordering of the first order RG equations and separation of the
coefficients of $e^{i\omega t}$ and $e^{-i\omega t}$, together with some
manipulation of the operator equations, lead us to the following closed set of
five coupled differential equations for five ordinary functions of $\tau$:
\begin{eqnarray}
\frac{1}{R}\frac{{\rm d} R}{{\rm d}\tau} & = & \frac{3 i \lambda
l^2}{4\omega}\left(\langle(\alpha^{\dag})^2\rangle e^{-2i\theta} -
\langle\alpha^2\rangle e^{2i\theta}\right)\nonumber\\
\frac{{\rm d}\theta}{{\rm d}\tau} & = & \frac{3\lambda}{8\omega}\left(R^2 + 4
l^2 + 2 l^2\left(\langle(\alpha^{\dag})^2\rangle e^{-2i\theta} +
\langle\alpha^2\rangle e^{2i\theta}\right) + 8 l^2\langle
N\rangle\right)\nonumber\\
\frac{{\rm d}\langle\alpha^2\rangle}{{\rm d}\tau} & = &\frac{-3 i \lambda
R^2}{8\omega}\left(4\langle\alpha^2\rangle + 2 e^{-2i\theta}\langle N\rangle +
e^{-2i\theta}\right)\nonumber\\
\frac{{\rm d}\langle\left(\alpha^{\dag}\right)^2\rangle}{{\rm d}\tau} & =
&\frac{3 i
\lambda R^2}{8\omega}\left(4\langle\left(\alpha^{\dag}\right)^2\rangle + 2
e^{2i\theta}\langle N\rangle + e^{2i\theta}\right)\nonumber\\
\frac{{\rm d}\langle N\rangle}{{\rm d}\tau} & = & \frac{-3 i
\lambda R^2}{8\omega}\left(\langle(\alpha^{\dag})^2\rangle e^{-2i\theta} -
\langle\alpha^2\rangle e^{2i\theta}\right)\label{conjunto}
\end{eqnarray}
These equations, after a little  reorganising in order to write them as
a set of real equations with no free parameters, can be solved numerically
without much further ado, leading for predictions for the long time behaviour
of $R(\tau)$ and $\theta(\tau)$.

Consider now that we constantly update $\tau$ to be $t$. The secular terms will
disappear in $\varphi_1$, and their effect will show up through $R$ and
$\theta$, now understood as functions of $t$:
\begin{equation}
\varphi= R(t) \cos(\omega t + \theta(t)) + \epsilon ({\rm r.t.})\,,
\end{equation}
where r.t. stands for regular terms, i.e., well controlled small perturbations
to the first term.

Moreover, it can be seen directly, from the analysis of equations
(\ref{conjunto}), that \begin{equation}
R\frac{{\rm d}R}{{\rm d}\tau} + 2 l^2\frac{{\rm d}\langle N\rangle}{{\rm
d}\tau}=0\,,
\end{equation} 
which is the equation that implements the conservation of energy
to this order.

The numerical solution for the initial conditions
$\varphi(0)=\sqrt{8\omega/(3\lambda)}$, $\langle N\rangle=0$, $\theta(0)=0$,
$\langle\alpha^2\rangle=0$ and $\langle\left(\alpha^{\dag}\right)^2\rangle=0$
is as shown in figure \ref{fig1}, where  $\varphi(t)$ is depicted. The units used
are arbitrary.

\section{Direct application of the RG method to the quantum Duffing equation}
In the previous section we have demonstrated the usefulness of the RG method in
eliminating the secular terms for the truncated equations obtained at one loop.
We shall now consider the whole quantum Duffing equation, eqn. (\ref{heiseq}),
and apply the same techniques for the complete problem. 

Let us perform a simple perturbation expansion in the $\lambda$ coupling
constant, $q=q_0+\lambda q_1 +O(\lambda^2)$. The solution to order 0 is
simply
\begin{equation}
q_0=\sqrt{\frac{\hbar}{2m\omega}}\left(\beta e^{-i\omega t} + \beta^{\dag}
e^{i\omega t}\right)\,.
\end{equation}
We have written $\beta$ and $\beta$ in order to make the distinction with the
operators $\alpha^{\dag}$ and $\alpha$, which were creation and annihilation
operators for fluctuations in the previous section, whereas $\beta^{\dag}$ and
$\beta$ are creation and annihilation operators for the full oscillator problem.
The first order equation presents resonance and, therefore, secular terms.
Writing down just the singular (i.e., secular) part of $q_1$, which we call
$q_{1,{\rm s}}$,
\begin{eqnarray}
q_{1,{\rm s}}&=&\left(\frac{\hbar}{2 m \omega}\right)^{3/2}
\frac{i(t-\tau)}{2\omega}\Big(\left(\beta\left(\beta^{\dag}\right)^2 +
\beta^{\dag}\beta\beta^{\dag} +
\left(\beta^{\dag}\right)^2\beta\right) e^{i\omega t} -\nonumber\\
&-& \left(\beta^2\beta^{\dag} +
\beta\beta^{\dag}\beta +
\beta^{\dag}\beta^2\right) e^{-i\omega t}\Big)\,.
\end{eqnarray}
Let us now allow a dependence in $\tau$ of the pair of operators $\beta$ and
$\beta^{\dag}$. Imposing the RG condition ${\rm d}q/{\rm d}\tau=0$, we obtain
\begin{equation}
\frac{{\rm d}\beta}{{\rm d}\tau} + \frac{i\lambda\hbar}{4m\omega^2}
\left(\beta^2\beta^{\dag} + \beta\beta^{\dag}\beta +
\beta^{\dag}\beta^2\right)=O(\lambda^2)\,,
\end{equation}
and the hermitian conjugate thereof. We notice that ${\cal
N}=\beta^{\dag}\beta$ and $[\beta,\beta^{\dag}]$ are constants under the flow
of $\tau$, which allows us to solve these equations in the form
\begin{eqnarray}
\beta(\tau) & = & \beta(0) e^{-3i\lambda\hbar{\cal
N}\tau/(4m\omega^2)}\,,\nonumber\\
\beta^{\dag}(\tau) & = &  e^{3i\lambda\hbar{\cal
N}\tau/(4m\omega^2)}\beta^{\dag}(0)\,,
\end{eqnarray}
which, on being substituted in the perturbative expansion of $q$, together with
the change $\tau\to t$, gives us
\begin{equation}
q(t) = \sqrt{\frac{\hbar}{2m\omega}}\left( e^{-i\omega t}\beta(0)
e^{-3i\lambda\hbar{\cal N}\tau/(4m\omega^2)} + e^{i\omega
t}e^{3i\lambda\hbar{\cal N}\tau/(4m\omega^2)}\beta^{\dag}(0)\right)\,.
\end{equation}
We have thus obtained an asymptotic expression for this operator. On computing
$\langle n-1|q(t)|n\rangle$, we see that the energy difference between levels
comes out as $E_n-E_{n-1}=\hbar\omega\left(1+(3\lambda\hbar
n)/(4m\omega^3)+O(\lambda^2)\right)$, consistent with all previous computations
of this quantity.

It has to be observed that our result is identical to the one obtained by
Bender and Bettencourt \cite{bb}, as is only to be expected, given the
equivalence of the multiple scales method and the RG methods for a wide class
of differential equations, to which the (classical) Duffing equation belongs. On
the other hand, note the simplicity of our approach, where no {\sl a priori}
scale has to be assumed.

In order to stress this latter point, let us consider the second order
computation for this problem. The source term for $q_2$ is given by $-(q_0^2
q_1 + q_0 q_1 q_0 + q_1 q_0^2)$, where we have to consider the full $q_1$ and
not just the singular part.  In this source term there will be terms that will
give rise to secularities of the form $e^{\pm 3 i \omega t}(t-\tau)$ and
$e^{\pm i\omega t}(t-\tau)^2$. These we shall be able to ignore, because the
renormalization to first order takes care of them. As a matter of fact, this is
precisely what the statement of perturbative renormalizability amounts to in
our case: that no divergences of a different form arise in the process of
renormalization, that is to say, that all divergent (secular) terms can be taken
care of by  renormalization of the terms $\beta e^{-i\omega t}$ and $\beta^{\dag}
e^{i\omega t}$. 

Another (simple) technicality in the problem at hand is that, since we have
checked that $[\beta,\beta^{\dag}]$ is constant to order $\lambda^2$, we can use
the commutator in the $\lambda$ and $\lambda^2$ terms, thus making the
computation somewhat easier.

This results in the following expression for the secular relevant part of
$q_2$, $q_{2,{\rm sr}}$:
\begin{equation}
q_{2,{\rm sr}} = \frac{-3il^5(t-\tau)}{16\omega^3}\left((5{\cal
N}^2-1) \beta^{\dag} e^{i\omega t}- \beta (5{\cal
N}^2-1) e^{-i\omega t}\right)\,,
\end{equation}
whence the improved RG equation reads
\begin{equation}
\frac{{\rm d}\beta}{{\rm d}\tau} + \frac{3i\lambda\hbar}{4m\omega^2}\beta{\cal
N} -\frac{3 i\lambda^2\hbar^2}{64m^2\omega^5}\beta(5{\cal
N}^2-1)=O(\lambda^3)\,,
\end{equation}
thus giving us 
\begin{equation}
\beta(\tau)  =  \beta(0) \exp\left(\frac{-3i\lambda\hbar{\cal
N}\tau}{(4m\omega^2)}+\frac{3 i\lambda^2 h^2(5{\cal N}^2-1)\tau}{64
m^2\omega^5}\right)\,,
\end{equation}
and, as a consequence, 
\begin{equation}
E_n-E_{n-1}=\hbar\omega\left(1+(3\lambda\hbar
n)/(4m\omega^3)-
3\lambda^2\hbar^2(5 n^2 -1)/(64 m^2\omega^6)
+O(\lambda^3)\right)\,.
\end{equation}

\section{Quantum parametric resonance}
As a last example of the usefulness of the RG method for quantum mechanical
problems, we shall now illustrate its application to the phenomenon of quantum
parametric resonance. Consider then the  following hamiltonian:
\begin{equation}
H=\frac{1}{2m}P^2 + \frac12 m\omega_0^2 (A + 2 q \cos(\omega_0 t)) X^2\,,
\end{equation}
where $A$ , $q$ and $\omega_0$ are constants.  The evolution of any given state
is computed by acting on it with the evolution operator $U(t,t_0)$,
which satisfies 
\begin{equation}
i\hbar\frac{\partial U}{\partial t}(t,t_0)=H(t) U(t,t_0)\,,
\end{equation}
with $U(t_0,t_0)=I$, the identity operator.

Let us divide the hamiltonian into an unperturbed and a perturbation part:
\begin{equation}
H=H_0 + H_1=\left(\frac{1}{2m}P^2 + \frac18 m\omega_0^2 X^2\right) +
\left(\frac12 m\omega_0^2 ((A-\frac14) + 2 q
\cos(\omega_0 t)) X^2\right)\,.
\end{equation}
The reason for this decomposition lies in our previous knowledge that resonance
will definitely set in if $A$ is equal to $1/4$, but this is not essential for
the final results.

The evolution operator can be written as 
\begin{equation}
U(t,t_0)=e^{-i(t-t_0) H_0 /\hbar} U_I(t,t_0)\,,
\end{equation}
in such a way that the interaction picture evolution operator obeys the
following equation:
\begin{equation}
i\hbar\frac{\partial U_I}{\partial t}(t,t_0)=H_I(t) U_I(t,t_0)\,,
\end{equation}
and in our case 
\begin{eqnarray}
H_I & = & \frac12 m\omega_0^2 ((A-\frac14) + 2 q
\cos(\omega_0 t)) X_I^2\nonumber\\
 & = &\frac{\hbar\omega_0}2((A-\frac14) + 2 q
\cos(\omega_0 t)) \left(e^{-i\omega_0 (t-t_0)/2}a +e^{i\omega_0
(t-t_0)/2}a^{\dag}\right)^2
\,.
\end{eqnarray}
The constant operators $a$ and $a^{\dag}$ are the annihilation and creation
operators at time $t_0$.

We now perform the usual perturbative expansion for the interaction picture
evolution operator, restricting ourselves to the Born formula,
$U_I=1-\frac{i}{\hbar}\int_{t_0}^t{\rm d}s\,H_I(s)$. However, this leads to
secular terms, and in order to eliminate them, we shall rather use this
approximation close to the time $t=\tau$, by using the initial condition
$U_I(\tau,t_0)=\alpha(\tau)$, such that
\begin{equation}
U_I(t,t_0)=\alpha(\tau)-\frac{i}{\hbar}\int_{\tau}^t{\rm
d}s\,H_I(s)\alpha(\tau)\,+{\rm~higher~order~terms}.
\end{equation}
Retaining only the secular terms, we obtain
\begin{equation}
U_I(t,t_0)=\alpha(\tau) -\frac{i\omega_0}2
(t-\tau)\left((A-\frac14)(aa^{\dag}+a^{\dag}a) + q (e^{i\omega_0 t_0}a^2 +
e^{-i\omega_0 t_0}(a^{\dag})^2)\right)\alpha(\tau)+{\rm r.~t.~}\,.
\end{equation}
For the sake of simplicity, let us set $t_0=0$, without any loss of generality.
We know that $U_I$ cannot depend on the choice of $\tau$, and we are thus
led to the RG equation to first order
\begin{equation}
\frac{\partial\alpha}{\partial\tau} +\frac{i\omega_0}2\left((A-\frac14)(aa^{\dag}
+a^{\dag}a) + q (a^2 + (a^{\dag})^2)\right)\alpha(\tau)=0\,,
\end{equation}
which, on being solved, provides us with an improved expression for the
interaction picture evolution operator, $U_I(t,0)=\exp(-i t H_{\rm eff}/\hbar)$,
where $H_{\rm eff}$ is the large time asymptotic effective hamiltonian read off
directly from the RG equation:
\begin{eqnarray}
H_{\rm eff} & = & \frac{\hbar\omega_0}2\left((A-\frac14)(aa^{\dag}+a^{\dag}a) 
+ q (a^2 + (a^{\dag})^2)\right)\nonumber\\
 & = & 
2\left(\frac1{2m} (A-\frac14-q) P^2 +\frac18 m\omega_0^2 (A-\frac14 +q)
X^2\right)\,.
\end{eqnarray}
This is the  first important result of our computation: we have resummed the
effect of the  variable frequency into an effective large time
constant hamiltonian. If it happens that, for small positive $q$, $\frac14-q<
A<\frac14+q$, this effective hamiltonian corresponds to an inverted oscillator,
thus marking the principal instability band (to the order we have computed).

It now behooves us to compute the creation of particles due to this
instability. In order to do this, we shall first write down the
integral kernel that corresponds to $U_I$ in the position representation,
$K(x,t;x',t')$, using standard results for quadratic hamiltonians \cite{FH}.
Let $\gamma=\omega_0\sqrt{q^2-(A-1/4)^2}$ and $\varphi =\sqrt{(q-1/4
+A)/(q+1/4-A)}$. The integral kernel $K(x,t;x',0):=\langle
x|U_I(t,0)|x'\rangle$ is computed to be (asymptotically):
\begin{equation}
K(x,t;x',0)=\left(\frac{i m \omega_0\varphi
}{4\pi\hbar\sinh(\gamma t)}\right)^{1/2}
\exp\left(\frac{-i m \omega_0\varphi}{4\hbar\sinh(\gamma t)}\left((x^2 +
x'^2)\cosh(\gamma t) - 2xx'\right)\right)\,.
\end{equation}
It is now feasible to compute the asymptotic value of  $\langle
n|U(t,0)|0\rangle$ through  simple tabulated integrals, and we can calculate the
transition probability from the ground state to even states:
\begin{equation}
P_{0\to2l}(t)=\frac{(2l)!}{(l!)^2 2^{2l}}
\frac{2\varphi}{\sqrt{4\varphi^2 +
(1+\varphi^2)^2 \sinh^2(\gamma
t)}}\left(\frac{(1+\varphi^2)^2\sinh^2(\gamma
t)}{4\varphi^2 +
(1+\varphi^2)^2\sinh^2(\gamma
t)}\right)^l\,.
\end{equation}
It is easy to check that unitarity is preserved. An analogous computation leads
to the rate of particle production, 
\begin{equation}
{\cal N}(t) = \frac{(1+\varphi^2)^2}{4\varphi^2}{\rm sinh}^2(\gamma t)\,.
\end{equation}
These results can be compared with the computations of Shtanov et al.
\cite{bts}, and coincide completely for the specific case at hand. Shtanov et
al. arrive to this result through Bogolyubov  transformations (to identify the
function giving particle creation) and Krylov-Bogolyubov averaging (to
perform the asymptotic analysis). This coincidence comes as no surprise given
the first order equivalence of Krylov-Bogolyubov averaging to two-timing (a
particular instance of the multiple-scales method) for a wide class of
differential equations \cite{smith}, and the (again first-order) equivalence of
the multiple-scales and  RG methods for many instances of equations.

\section{Conclusions}
We have performed several operator computations in quantum mechanics using the
RG method for global asymptotic analysis. This method has the serious advantage
that unitarity is built in, and that computations are simpler and more direct
than in other techniques for asymptotic analysis. These ideas should be useful
in a wide realm of applications. In the special case of quantum parametric
resonance we derive explicitly an (asymptotic) effective hamiltonian, which is
an inverted harmonic oscillator whenever the system is in the instability
region: the instability associated with the parametric resonance is turned into
the unboundedness of the interaction hamiltonian, thus demonstrating the basic
equivalence (asymptotically) of such different systems. Even so, unitarity is
preserved throughout our computation, and the asymptotic results we obtain are
well-behaved with respect to this fundamental property of quantum mechanical
evolution. We have  performed an analysis of the first instability
band only for the quantum Mathieu equation. However, it is
possible within this method to examine the whole forbidden/allowed band
structure of this model, following in the quantum context the study carried out
for the classical case by Goldenfeld and collaborators. As a matter of fact, for
quadratic hamiltonians the whole instability analysis can be reduced to
classical mechanics, i.e., to classical Mathieu (or similar) equations. What we
emphasise as novel in our results is the interpretation of instabilities as
being due to effective large time inverted oscillator hamiltonians. Furthermore,
analogous analyses can be carried for non-quadratic hamiltonians, even time
dependent, where the quantum mechanical character of the problem would show
itself to its fullest.

\begin{figure}
\caption{Backreaction: $\varphi(t)$.} \label{fig1}
\end{figure}

 
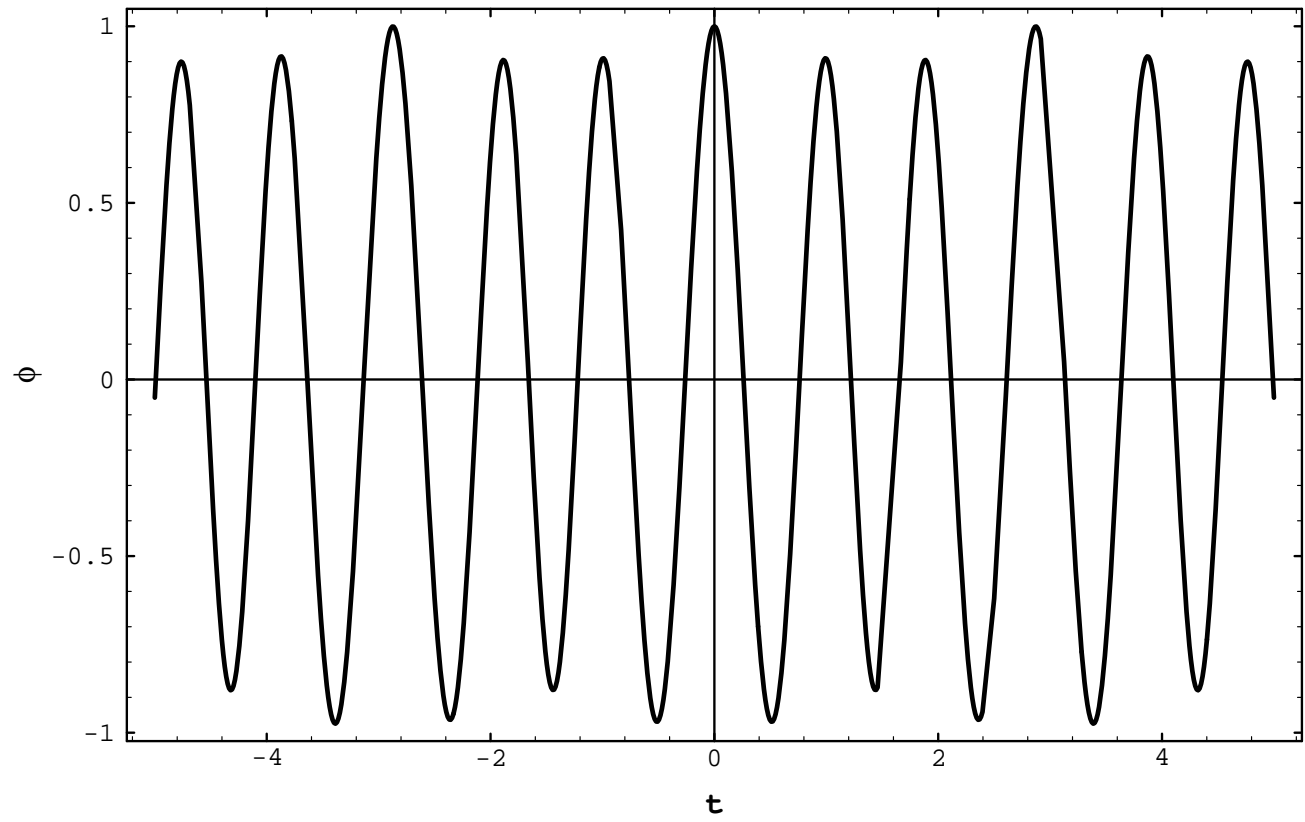
\end{document}